\documentclass[10pt,twoside]{article}
\usepackage{graphicx}
\makeindex

\begin{document}

\title{Accelerating multidimensional cosmologies with scalar fields}

\markboth{Baukh Victor}{Accelerating multidimensional cosmologies
with scalar fields}

\author{Baukh Victor
\\[5mm]
\it Department of Theoretical Physics,  Odessa National University\\
 Street Dvoryanskaya, 2, 65026 Odessa, Ukraine\\
\it e-mail: bauch\_vgr@ukr.net }

\date{}
\maketitle

\thispagestyle{empty}

\begin{abstract}
\noindent We study multidimensional cosmological models with a
higher-dimensional product manifold
 $R \times M_0 \times M_1$ (where $M_0$ and $M_1$ are spherical and
 flat spaces, correspondingly) in the presence of a minimal free scalar field.
 The dimensions of spaces $M_i\, (i=0,1)$ are not fixed
 and the choice of an external (our) space in the product manifold is arbitrary.
 Dynamical behaviour of the model is analyzed both in Einstein and Brans-Dicke
 conformal
 frames. For a number of particular cases, it is shown that external space-time
 undergoes
 an accelerated expansion.\\

\end{abstract}

\section{Introduction}

Distance measurements of type Ia supernovae (SNe,Ia)
\cite{supernova} as well as cosmic microwave background (CMB)
anisotropy measurements \cite{cmb1} performed during the last
years give strong evidence for the existence of dark energy --- a
smooth energy density with negative pressure which causes an
accelerated expansion of the Universe at present time.
The challenge to theoretical cosmology consists in finding a
natural explanation of
such acceleration (dark energy).

In the present paper we consider a multidimensional cosmological
model with a higher-dimensional product manifold which consists of
spherical and flat spaces with dimensions $d_0$ and $d_1$
correspondingly in the presence of a minimally coupled free scalar
field. We investigate this model to find under which circumstances
the external spacetime undergoes an accelerated expansion with
simultaneous compactification of the internal space.

To start with, let us consider a cosmological model with
factorizable geometry
\begin{equation} g= -\exp[2\gamma(\tau)]d\tau\otimes d\tau +
a_0^2(\tau) g_0 + a_1^2(\tau) g_1\, , \end{equation} which is
defined on the manifold
\begin{equation}
M = R \times M_0 \times M_1\,
\end{equation}
and with minimal free scalar field $\varphi$. We assume that
factors $M_i$ are Einstein spaces: $R[g_0]=R_0
> 0,\; R[g_1]=0$. The action for considered model reads
\begin{equation}
S=\frac{1}{2\kappa^2}\int\limits_M d^Dx\sqrt{|g|}{R} +
\int\limits_M
d^Dx\sqrt{|g|}\left[-\frac{1}{2}e^{-2\gamma(\tau)}\partial_M\varphi\partial_N\varphi\right]
+ S_{YGH}\, , \end{equation} where $S_{YGH}$ is the
York-Gibbons-Hawking boundary term.

It can be easily seen \cite{BZ} that in the harmonic time gauge
$\gamma=\gamma_0$ (see \cite{IMZ}) equations of motion have the
following solutions:
\begin{eqnarray}
\nonumber
a_0(\tau)&=&a_{(c)0}\frac{\exp{\left(-\frac{1}{d_0-1}\sqrt{\frac{d_1(d_0-1)}{D-2}}p^1
\tau\right)}}
{\left[\cosh{(\sqrt{\frac{(d_0-1)2\varepsilon}{d_0}}\tau)}\right]^{\frac{1}{d_0-1}}}\, ,\\
\label{solution}
a_1(\tau)&=&a_{(c)1}\exp\left({\frac{1}{d_1}\sqrt{\frac{d_1(d_0-1)}{D-2}}p^1
\tau}\right),\quad\\
 \varphi(\tau)&=&p^2\tau+q^2\, ,\nonumber
\end{eqnarray}
where $a_{(c)0,1},\, p^{1,2}$ and $q^2$ are constants of
integrations and $2\epsilon = {(p^1)}^2 + {(p^2)}^2 >0$.

We shall analyze a dynamical behaviour of the model both in
Brans-Dicke and Einstein conformal frames where metrics are
connected as follows (\cite{RZ})
\begin{eqnarray}
 \nonumber g&=&-e^{2\gamma} d\tau\otimes d\tau + a_e^2
g^{(e)} + a_i^2 g^{(i)} \\\label{metrika2}&=& -dt\otimes dt +
a_e^2 g^{(e)} + a_i^2 g^{(i)} \\\nonumber&=&
\Omega^2(-d\tilde{t}\otimes d\tilde{t} + \tilde{a}_e^2 g^{(e)}) +
a_i^2 g^{(i)}\, . \end{eqnarray} Here, $t$ ($\tilde{t}$) is a
physical/synchronous time
and $a_e, a_i$ ($\tilde{a}_e, \tilde{a}_i$) are scale-factors of
the external and internal spaces in the Brans-Dicke (Einstein)
frame. Conformal factor $\Omega = a_i^{-d_i}$ (in the case of 4-D
external spacetime).


\section{Dynamical behaviour in the Brans-Dicke frame}

Here, the Hubble and the deceleration parameters can be expressed
via harmonic time as follows:
$$
H_0=\frac{1}{a_0}\frac{da_0}{dt} =-\frac{\xi_1+\xi_2
\tanh(\xi_2\tau)}{f(\tau)(d_0-1)},\quad -q_0
=\frac{1}{a_0}\frac{d^2a_0}{dt^2}
=-\xi_2\frac{\xi_2+\xi_1\tanh(\xi_2\tau)}{f^2(\tau)(d_0-1)}\, ,
$$
$$
H_1=\frac{1}{a_1}\frac{da_1}{dt}=\frac{\xi_1}{f(\tau)d_1},\quad
-q_1=\frac{1}{a_1}\frac{d^2a_1}{dt^2}=\frac{\xi_1((D-2)\xi_1+d_0d_1\xi_2
\tanh(\xi_2\tau))}{f^2(\tau)d_1^2(d_0-1)},
$$
where $f(\tau)=\frac{dt}{d\tau}$, $\xi_1=
\sqrt{\frac{d_1(d_0-1)}{D-2}}p^1$ ,\quad $\xi_2=\sqrt
{\frac{(d_0-1)2\varepsilon} {d_0}},$\quad . We consider the
expanding space as an external one. Thus, we arrived to two
separate cases:

{\bf{i. $M_0$ is the external space ($\xi_1<0$):}}

In this case, we find that $M_0$ undergoes an accelerated
expansion only if $|\xi_1|>\xi_2$ with simultaneous
compactification of $M_1$. For the asymptotical behaviour $\tau
\to \infty$ we have 2 particular cases where external space
undergoes accelerated expansion:

1.{\it $d_0\xi_2>|\xi_1|>\xi_2$, Big Rip scenario}:

In this case the scale-factor of $M_0$ is
\begin{equation} a_0(t)\simeq a_{(c)0}\left[\frac{
(d_0\xi_2-|\xi_1|)(c-t)}{(d_0-1) 2^{\frac {d_0}{d_0-1}}
a_{(c)0}^{d_0} a_{(c)1}^{d_1}}\right]^{-\frac{|\xi_1| - \xi_2}{
d_0 \xi_2-|\xi_1|}}\, . \end{equation} It can be easily seen that
the Universe behaves according to the Big Rip scenario, i.e. the
scale-factor $a_0$ achieves infinite value in the finite period of
time.

2.{\it{$|\xi_1|>d_0\xi_2$}}:

Here, for the scale factor of the external space we obtain the
asymptote:
\begin{equation} a_0(t)\simeq a_{(c)0}\left[ \frac{ (|\xi_1| - d_0
\xi_2)(t-c)}{(d_0-1) 2^{\frac {d_0}{d_0-1}} a_{(c)0}^{d_0}
a_{(c)1}^{d_1}}\right]^{\frac{|\xi_1| - \xi_2}{|\xi_1| - d_0
\xi_2}}\, .\end{equation}

For these cases, the following relations between momentum $p^1$
and kinetic energy $\varepsilon$ hold:
\begin{itemize}
\item case $d_0\xi_2>|\xi_1|>\xi_2$:\quad $\frac{d_0d_1
(p^1)^2}{(D-2)}2\varepsilon > \frac{d_1(p^1)^2}{d_0(D-2)}$, i.e
real scalar field and Big Rip scenario

 \item
case $|\xi_1|>d_0\xi_2$: \quad $2\varepsilon < \frac{d_1
(p^1)^2}{d_0(D-2)}$, so $2\varepsilon<{(p^1)}^2$, i.e it is
possible only if $(p^2)^2=(\dot\varphi(\tau))^2<0$  and scalar
field is imaginary
\end{itemize}

{\bf{ii. $M_1$ is the external space ($\xi_1>0$):}}

In this case space $M_1$ undergoes an accelerating expansion (via
{\it Big Rip scenario}) with simultaneous compactification of
$M_0$. Such behaviour for $M_1$ can be easily traced from
asymptotical behaviour of $a_1(t)$:
\begin{equation} a_1(t)\simeq a_{(c)1}\left[ \frac{ (\xi_1 + d_0
\xi_2)(c-t)}{(d_0-1) 2^{\frac {d_0}{d_0-1}} a_{(c)0}^{d_0}
a_{(c)1}^{d_1}}\right]^{-\frac{\xi_1 (d_0-1)}{(\xi_1 + d_0
\xi_2)d_1}}\, .
\end{equation}

\section{Dynamical behaviour in the Einstein frame}
Now, we analyze the behaviour of the model in the Einstein frame.
Here, we also consider two separate cases:

{{\bf{i. $M_0$ is the external space:}}

If $M_0$ is the external space, its deceleration parameter is

\begin{equation}
-\tilde{q}_0 = \frac{1}{\tilde a_0 } \frac{d^2\tilde a_0}{d \tilde
t^2 } =-\frac{\xi_2^2}{f_{E0}^2(\tau)(d_0-1)} < 0\, ,
\end{equation}
where $f_{E0}(\tau)=\frac{d\tilde{t}}{d\tau}$. It can be easily
seen that in this case an accelerated expansion of the external
space is impossible.

{\bf{ii. $M_1$ is the external space:}}

If $M_1$ is the external space, its deceleration parameter reads:
 \begin{eqnarray}
 \label{accel1e}
-\tilde{q}_1 &=& \frac{1}{\tilde a_1 } \frac{d^2\tilde a_1}{d
\tilde t^2 }=-\frac{1}{f_{E1}^2(\tau)d_1^2(d_0-1)^2(d_1-1)}\times
\\ \nonumber &\times&
\left[[(D-2)\xi_1+d_1d_0\xi_2\tanh(\xi_2\tau)]^2 +
\frac{d_0d_1^2(d_0-1)\xi_2^2}{\cosh^2(\xi_2\tau)}\right]< 0 \,
,\end{eqnarray} where $f_{E1}(\tau)=\frac{d\tilde{t}}{d\tau}$ .
Accelerated expansion of the external space is impossible in this
case too.

\section*{Summary}
Briefly, we can summarize our results as follows:

{\bf {In the Brans-Dicke frame}}:

i. accelerated expansion of the positive curvature space can be
archived either if it undergoes the Big Rip scenario or it
contains scalar field with negative kinetic term.

ii. Ricci-flat space undergoes late time acceleration via Big Rip
scenario and its qualitative behavior does not depends on a
particular choice of parameters of the model.

{\bf {In the Einstein frame}}:

it is impossible to achieve an accelerated expansion of the
external space (either $M_0$ or $M_1$).

\section*{Acknowledgements}

The author is thankful to Alexander Zhuk for guidance and
assistance.


\end{document}